\documentclass[a4paper,10pt,reqno]{amsart}
\usepackage[margin=3cm]{geometry}
\usepackage{graphicx}
\usepackage{dsfont,amsmath,amssymb,color,units,pstricks}

\usepackage[first,bottomafter,timestamp]{draftcopy}

\newcommand \be  {\begin{equation}}
\newcommand \bea {\begin{eqnarray} \nonumber }
\newcommand \ee  {\end{equation}}
\newcommand \eea {\end{eqnarray}}

\title{Individual and collective stock dynamics: intra-day seasonalities}
\author{Romain Allez \and Jean-Philippe~Bouchaud} 
\address{Capital~Fund~Management, 6--8 boulevard Haussmann, 75009 Paris, France}
\email{Romain.Allez@cfm.fr, Jean-Philippe.Bouchaud@cfm.fr}
\urladdr{http://www.cfm.fr}
\date{\today}

\begin{document}

\begin{abstract}
We establish several new stylised facts concerning the intra-day seasonalities of stock dynamics. Beyond the well known U-shaped
pattern of the volatility, we find that the average correlation between stocks increases throughout the day, leading to a smaller relative dispersion
between stocks. Somewhat paradoxically, the kurtosis (a measure of volatility surprises) reaches a minimum at the open of the market, when the volatility is at its peak. 
We confirm that the dispersion kurtosis is a markedly decreasing function of the index return. This
means that during large market swings, the idiosyncratic component of the stock dynamics becomes sub-dominant. In a nutshell, early 
hours of trading are dominated by idiosyncratic or sector specific effects with little surprises, whereas the influence of the market factor 
increases throughout the day, and surprises become more frequent.
\end{abstract}

\maketitle
\section{Introduction}

Financial markets operate in sync with human activities. It is therefore no surprise that financial time series reveal a number of seasonalities 
related to human rhythms: markets open in the morning and close in the evening, remain closed during week-ends and during vacations; wages are
paid and portfolios are re-balanced on a monthly basis, earnings are announced on a quarterly basis (in the US), etc. These periodicities leave a
statistical trace on the time series of returns of many assets. Among the best known periodicities is the so-called U effect \cite{admati88,andersen97}, that describes the
intra-day pattern of volatility of individual US stocks: the average volatility is observed to be high after the market opens, then decreases as to reach a
minimum around lunch time and increases again steadily until market close.\footnote{This pattern is a little different in Europe or in the UK, with a 
second volatility spike at 2:30 pm GMT when the US market opens.}

In this short note we want to report on additional intra-day patterns concerning both individual and collective stock dynamics. First, we study the
intra-day pattern of other moments of the individual stock dynamics, beyond the well known U-shaped volatility. Second, we characterise the cross-sectional 
distribution of returns and its typical evolution during the day. Finally, we study the correlation matrix between stock returns and find that the
leading modes also have a very well defined intra-day pattern. Our study here is entirely empirical, but our results certainly beg for a 
detailed theoretical interpretation in terms of agent behaviour: strategies, information processing, risk aversion, etc. 
We provide some hints in that direction in the conclusion.

\section{Data, notations and definitions} 

We have considered a set of $N=126$ stocks of the New York Stock Exchange 
(which are among the $250$ largest market capitalisations) that has been continuously traded during the period between $01/01/2000$ and $12/31/2009$ to 
form a statistical ensemble of $5$ minutes stock returns. The total number of $5$ minute bins is $186,498$, whereas the total number
of days is $T=2,391$ ($K=78$ bins per day). Stocks will be labelled by $\alpha=1,\dots,N$, days by $t=1, \dots, T$ and bins by $k=1, \dots, K$. The
return of stock $\alpha$ in bin $k$ of day $t$ will be denoted as $\eta_\alpha(k;t)$. 

Different types of averages will be needed. Time averages for a given stock and a given bin are expressed with angled brackets: $\langle \dots \rangle$, 
whereas averages over the ensemble of stocks for a given bin in a given day appear with square brackets: $[ \dots ]$. For an arbitrary function $F(.)$ 
of these returns we therefore write:
\be
\langle F \rangle(k;\alpha)  := \frac{1}{T} \sum_{t=1}^T F(\eta_\alpha(k;t));\quad
[ F ](k;t) := \frac{1}{N} \sum_{\alpha=1}^N F(\eta_\alpha(k;t));\quad
[\langle F \rangle](k) =  \frac{1}{T} \sum_{t=1}^T [ F ](k;t).
\ee

The first set of observables
concerns single stock properties. We characterise the distribution of stock $\alpha$ in bin $k$ by its four first moments: mean $\mu_\alpha(k)$, 
standard deviation (volatility) $\sigma_\alpha(k)$, skewness $\zeta_\alpha(k)$ and kurtosis $\kappa_\alpha(k)$. We will in fact 
use low moments, less noisy estimates of the last two quantities. We will define $m_\alpha(k)$ as the median of all returns of stock $\alpha$ 
in bin $k$, and define:
\begin{subequations}\label{single}
\begin{align}
	\mu_\alpha(k)&= \langle \eta_\alpha(k;t) \rangle\\
	\sigma_\alpha^2(k)&=\langle \eta_\alpha(k;t)^2 \rangle - \mu_\alpha^2(k)\\
	\zeta_\alpha(k)&=\frac{6}{\sigma_\alpha(k)}\left(\mu_\alpha(k)-m_\alpha(k)\right)\\
	\kappa_\alpha(k)&=24 \left(1 - \sqrt{\frac{\pi}{2}} \frac{\langle |\eta_\alpha(k;t) -  \mu_\alpha(k)|}{\sigma_\alpha(k)} \rangle\right) + \zeta_\alpha(k)^2.\label{kurtosis}
\end{align}
\end{subequations}
Within a cumulant expansion, the last two lines coincide with the usual definition of skewness and kurtosis, but no moments larger than two are needed to
estimate them. Note that the correction term $\zeta_\alpha(k)^2$ to the kurtosis turns out to be negligible, and we have neglected it in the following.  
We will be interested below in the average over all stocks of the above quantities, as a way to characterise the typical intra-day evolution of the 
distribution of single stock returns.

One can also consider {\it cross sectional} distributions, i.e. the dispersion of the returns of the $N$ stocks for a given bin $k$ in a given day $t$, 
i.e. one distribution every five minutes. One can again characterise these distributions in terms of the first four moments. The median of all $N$ 
returns for a given $k;t$ is now $m_d(k;t)$ ($d$ for ``dispersion''), and we define:
\begin{subequations}\label{collective}
\begin{align}
	\mu_d(k;t)&= [\eta_\alpha(k;t)]\\
	\sigma_d^2(k;t)&=[ \eta_\alpha(k;t)^2 ] - \mu_d^2(k;t)\\
	\zeta_d(k;t)&=\frac{6}{\sigma_d(k;t)}\left(\mu_d(k;t)-m_d(k;t)\right)\\
	\kappa_d(k)&=24 \left(1 - \sqrt{\frac{\pi}{2}} \frac{[ |\eta_\alpha(k;t) -  \mu_\alpha(k)| ]}{\sigma_d(k)}\right)
\end{align}
\end{subequations}
Note that $\mu_d(k;t)$ can be seen as the return of an index, equiweighted on all stocks. We will be interested below in the average over all days 
of the above quantities, as a way to characterise the typical intra-day evolution of the dispersion between stock returns. 

Although the dispersion already captures part of the ``Co-movements'' of stocks, a more direct characterisation is through the standard correlation of returns.
In order to measure the correlation matrix of the returns, we first normalise each return by the dispersion of the corresponding bin. This factors in
any ``trivial'' intra-day seasonality, and also accounts for the fact that the volatility fluctuates quite a bit during the 10 year time interval that 
we consider. Therefore, we introduce: $\widehat \eta_\alpha(k;t)=\eta_\alpha(k;t)/\sigma_d(k;t)$ and study the correlation matrix defined for a 
given bin $k$:
\be\label{correlation}
C_{\alpha,\beta}(k) := \frac{1}{\widehat \sigma_\alpha(k)\widehat \sigma_\beta(k)} \langle \widehat \eta_\alpha(k;t) \widehat \eta_\beta(k;t) \rangle_c,
\ee
where the subscript $c$ means ``connected part'' (i.e. averages have been subtracted) and $\sigma_\alpha^2(k) := \langle \eta_\alpha^2(k;t) \rangle_c$.
Of special interest are the largest eigenvalues and eigenvectors of $C_{\alpha,\beta}(k)$, which charaterize the correlation structure of stock returns.
This analysis has been performed in several papers (\cite{bouchaudpotters04,BPshortreview,Utsugi,Plerou}) using daily or high frequency returns, and it is well known that the structure of large eigenvectors 
reflects the existence of economic sectors of activity. The largest eigenvalue $\lambda_1$, in particular, corresponds to the market mode, and is 
associated to an eigenvector with all entries positive and close to $1/\sqrt{N}$. In fact, $\lambda_1/N$ can be seen as a measure of the average 
correlation between stocks. We will be interested below in the $k$ dependence of the largest eigenvalues and their associated eigenvectors, a study that,
to the best of our knowledge, has not been reported in the literature before. 

\section{Single stock intra-day seasonalities}

\begin{figure} 
		\center
		\includegraphics[scale=0.45]{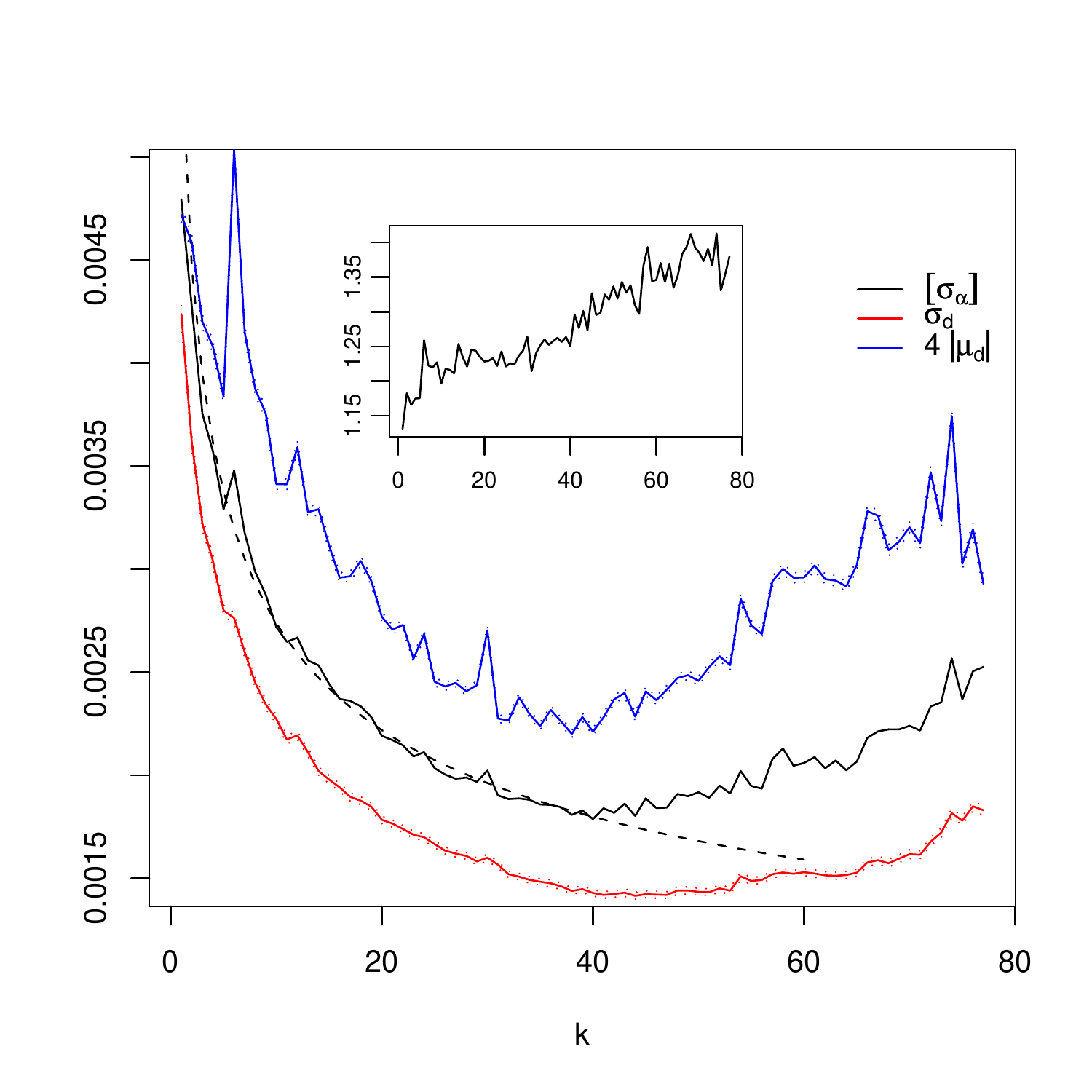}
		\caption{We show the average volatility of stocks $\sigma(k)$, the average cross sectional dispersion $\sigma_d(k)$ and 
		the average absolute value of the index return $\langle |\mu_d(k,t)|\rangle$ (multiplied by $4$ for clarity) as a function of $k$, with the corresponding statistical error bars. 
		All display the well known U pattern. We also compare $\sigma(k)$ with a power-law decay $k^{-\beta}$ with $\beta \approx 0.3$ (dashed line), which is a 
		good fit for the first half of the day. Inset: ratio $\sigma(k) / \sigma_d(k)$ as a function of $k$, showing that dispersion effects diminish throughout the day.} 
        \label{figure1}
\end{figure}

\subsection{Odd moments}

Odd moments tend to be small and noisy, so it is difficult to draw definite conclusions. The average return is on average over the whole period positive, 
but noisy and does not show any intra-day pattern. The average skewness of five minutes returns is also noisy and is compatible with zero, again without any 
identifiable intra-day pattern at all. This is at variance with the skewness of returns on a longer time 
interval, which is negative. The build up of negative skewness with time scale is a consequence of the leverage effect, i.e. negative returns tend to
be followed by larger volatilities (see e.g. \cite{bouchaudpotters04}). 

\subsection{Even moments}

The average volatility, on the other hand, reveals a very clean U-shaped pattern that has been reported many times in the literature (\cite{admati88,andersen97}). We show in Fig.~\ref{figure1} 
$\sigma(k)=[\sigma_\alpha(k)]$. Note that the overnight volatility $\approx 1.15 \%$ is much larger than the 
typical five minute volatility, and is not shown in the graph. Interestingly, the average volatility is found to decay in the first two hours of trading 
as a power-law $k^{-\beta}$ with $\beta \approx 0.3$. This relaxation is reminiscent of the power-law decay of the volatility after large price swings \cite{Omori,Muzy,Kertesz,Joulin}. 
The overnight return is indeed usually quite large, and can be seen as a strong perturbation. 
The power-law relaxation suggests that some critical mechanism is involved in the way volatility reverts back to `normal' after market jumps.

\begin{figure} 
		\center
		\includegraphics[scale=0.45]{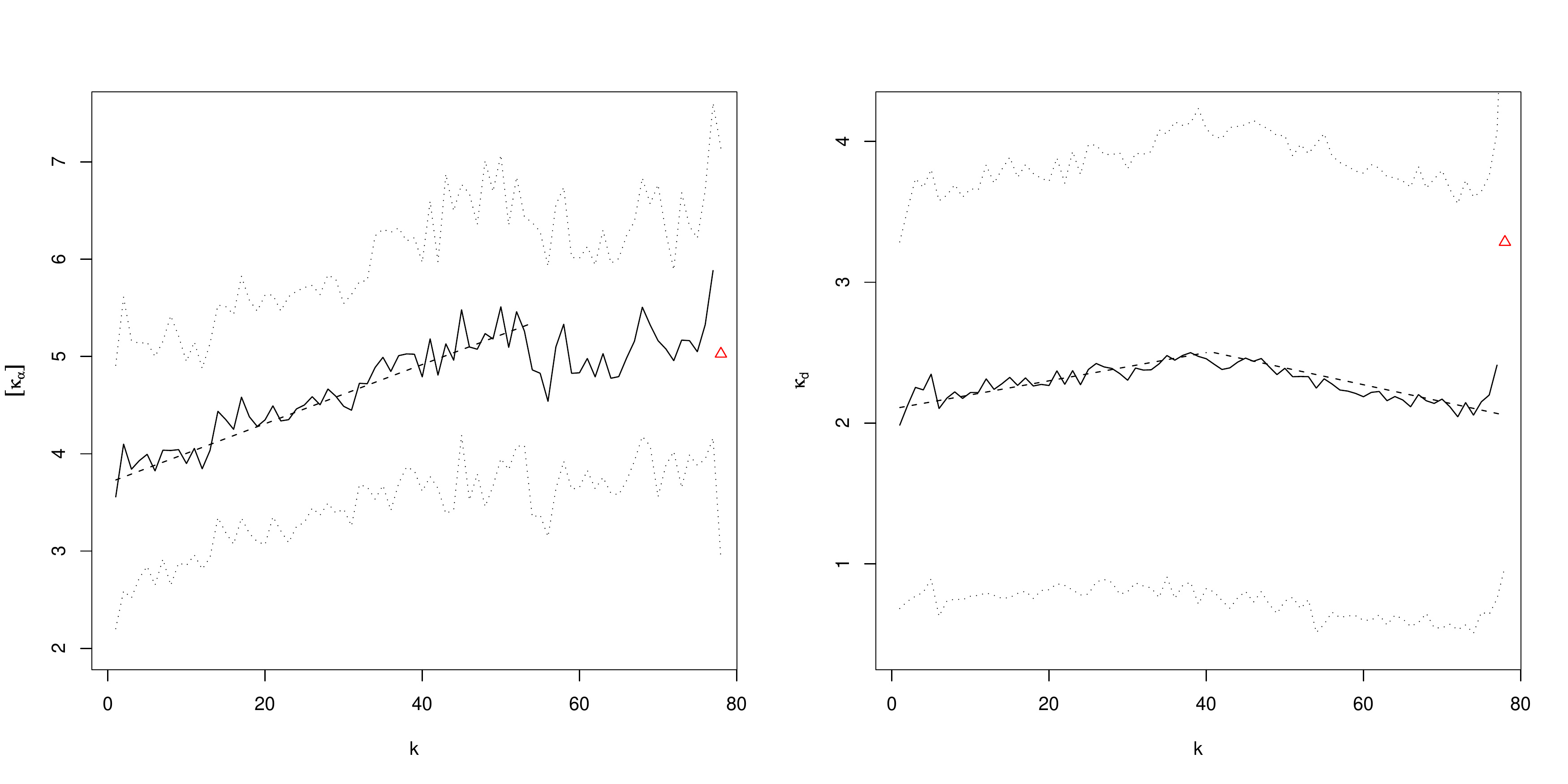}
		\caption{Left: average kurtosis of individual stocks $\kappa(k)$  as a function of $k$. Right:  average cross sectional kurtosis $\kappa_d(k)$ as a function of $k$.
        In both cases, we show the 1-$\sigma$ dispersion around the mean (i.e. not the error bar). The red triangles correspond to the overnight values.} 
        \label{figure2}
\end{figure}

Turning now to the kurtosis $\kappa(k)=[\kappa_\alpha(k)]$, we find, perhaps surprisingly, that there is a clear intra-day {\it growth} of the kurtosis
from $\kappa \approx 3.5$ at the beginning of the day to $\kappa \approx 5$ around 1 p.m., and stays approximately constant (but noisy) until the end 
of the day -- see Fig. \ref{figure2}, left. The overnight kurtosis remains around $5$. This finding is counter-intuitive because one would naively associate the large volatility 
in the morning with huge swings, symptomatic of the market uncertainty at the open. But this is not the case: the maximum of the intra-day volatility 
corresponds to a {\it minimum} in kurtosis. We will report similar counter-intuititive
results below. Possible mechanisms are discussed in the conclusion.

\section{Cross-sectional intra-day seasonalities} 

\subsection{Odd moments}

Noting that the average over stocks of $\mu_\alpha(k)$ is identical to the average over time of $\mu_d(k;t)$, the discussion of the first moment of
the cross-sectional distribution is redundant. The average of $|\mu_d(k;t)|$ is a proxy for the index volatility, and is displayed in Fig.~\ref{figure1} : it shows a
U-shaped pattern similar to that of $\sigma(k)$, with however a stronger end-of-day surge. This is due to the correlation pattern discussed in section 5 
below: the average correlation between stock indeed increases as the day proceeds, leading to an increased index volatility.

As far as the average skewness $\zeta_d(k)=\langle \zeta_d(k;t) \rangle$ is concerned, we again find a very
noisy quantity with no particular intra-day pattern. The only notable feature is that this time, the skewness is significantly positive, albeit small:
the average over $k$ of $\zeta_d(k)$ is found to be $\approx 0.025$. 

\subsection{Even moments}

As above, the even moments show clear patterns. The average dispersion $\sigma_d(k)=\langle \sigma_d(k;t) \rangle$ exhibits a U-shaped pattern 
very similar to that of $\sigma(k)$ --- see Fig.~\ref{figure1}. In fact, the ratio $\sigma(k)/\sigma_d(k)$ is plotted in the inset of Fig.~\ref{figure1} as a function of $k$ 
and increases from $\approx 1.15$ at the open to $1.45$ at the close. In relative terms, the dispersion is thus stronger in the morning, and 
decreases as the day proceeds.

The dispersion kurtosis $\kappa_d(k)=\langle \kappa_d(k;t) \rangle$, on the other hand, has an {\it inverted} U shape, and
reaches a minimum at the open and at the close of the market, i.e. when the dispersion and the volatility are locally maximum. So even when the 
dispersion of returns is at its peak, with stocks all over the place (so to say), the cross-sectional distribution of returns is on average 
closer to a Gaussian! Note however that the variation of the kurtosis is not large, from $\kappa_d=2$ to $\kappa_d=2.4$. 
The overnight dispersion kurtosis, on the other hand, is much stronger: $\kappa_d \approx 3.3$.

\subsection{Conditioning on the index return}

\begin{figure} 
		\center
		\includegraphics[scale=0.45]{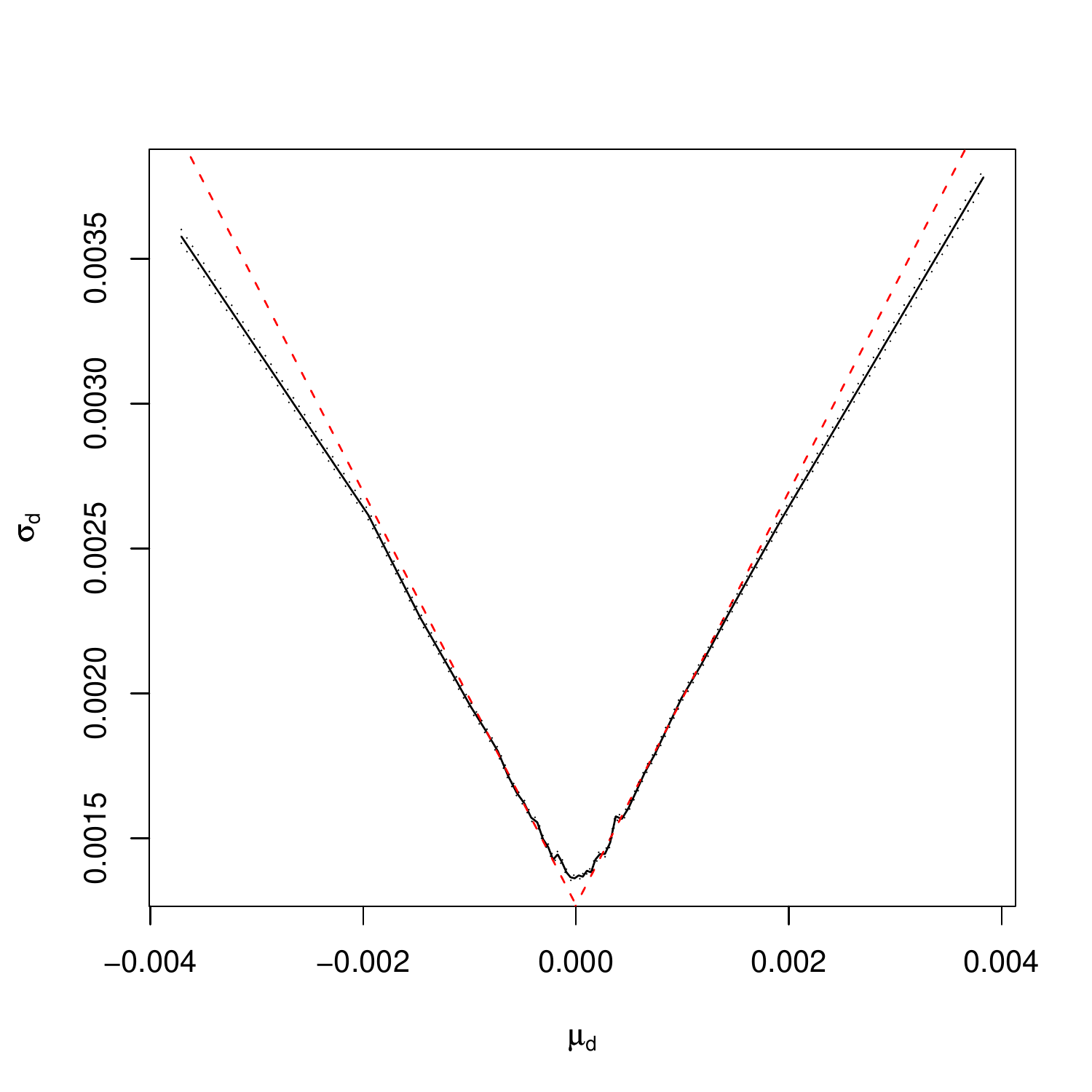}
		\caption{Cross sectional dispersion $\sigma_d$ as a function of the index return (equiweighted on all stocks) $\mu_d$. We added error bars, and linear branches 
		that fit the small $|\mu_d|$ slopes, that emphasise the sub-linear behaviour of $\sigma_d$.} 
        \label{figure3}
\end{figure}

As noted above, the quantity $\mu_d(k;t)$ is the return of an equiweighted index. It is interesting to condition the value of the moments of the 
cross-sectional dispersion on this quantity. Such a study was performed on daily returns in \cite{LilloMantegna} and more recently by L. Borland
\cite{Lisa}. In agreement with the results of \cite{LilloMantegna,Cizeau}, we find that the average dispersion $\sigma_d$ is an increasing function 
of the amplitude of the index return, see Fig.~\ref{figure3}. As noted in \cite{Cizeau}, this observation shows that the volatility of the stock 
residuals in a one-factor model must depend on the volatility of the market mode. Fig.~\ref{figure3} furthermore suggests that this dependence is sub-linear (see
\cite{Remi} for some elaborations on this observation.)

\begin{figure} 
		\center
		\includegraphics[scale=0.45]{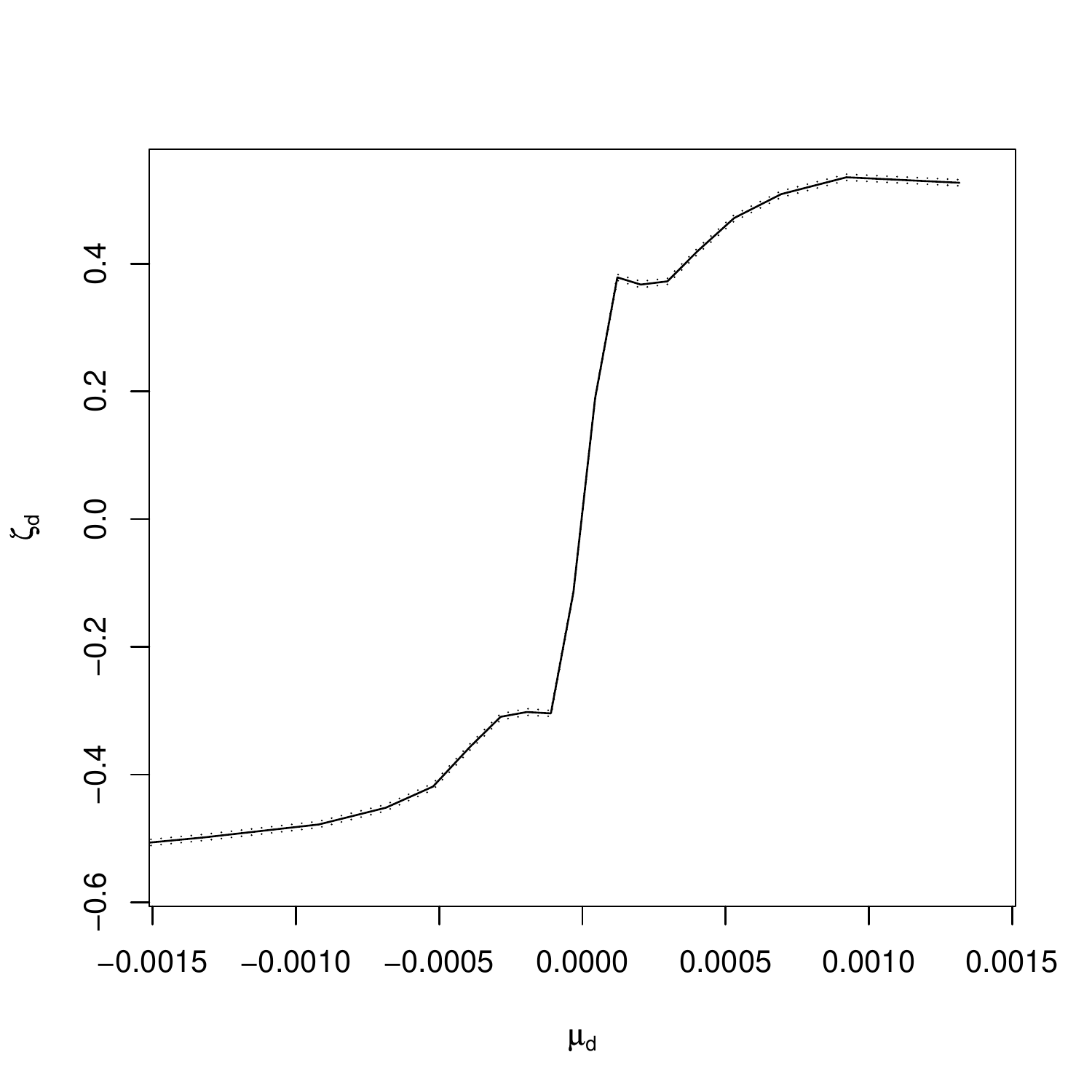}
		\caption{Cross sectional skewness $\zeta_d$ as a function of the index return (equiweighted on all stocks) $\mu_d$. We added error bars (dotted lines), 
		that are actually difficult to see near the origin.} 
        \label{figure4}
\end{figure}

As first established in \cite{Lillo} on daily data, we find that the skewness $\zeta_d$ is an odd function of $\mu_d$, as shown in 
Fig.~\ref{figure4}. Note that the skewness increases very abruptly for small $\mu_d$ and saturates for larger values of the index return. Pictorially, a positive
index return can be thought of as resulting from a few ``winners'' running ahead of the pack, contributing both to the mean $\mu_d$ and to the skewness.
The slope of $\zeta_d(\mu_d)$ around the origin, together with the fact that the index has made on average positive daily gains in the 
period 2000 -- 2009, are enough to explain the average value of the dispersion skewness $\zeta_d(k) \approx 0.025$ reported above. 

\begin{figure} 
		\center
		\includegraphics[scale=0.45]{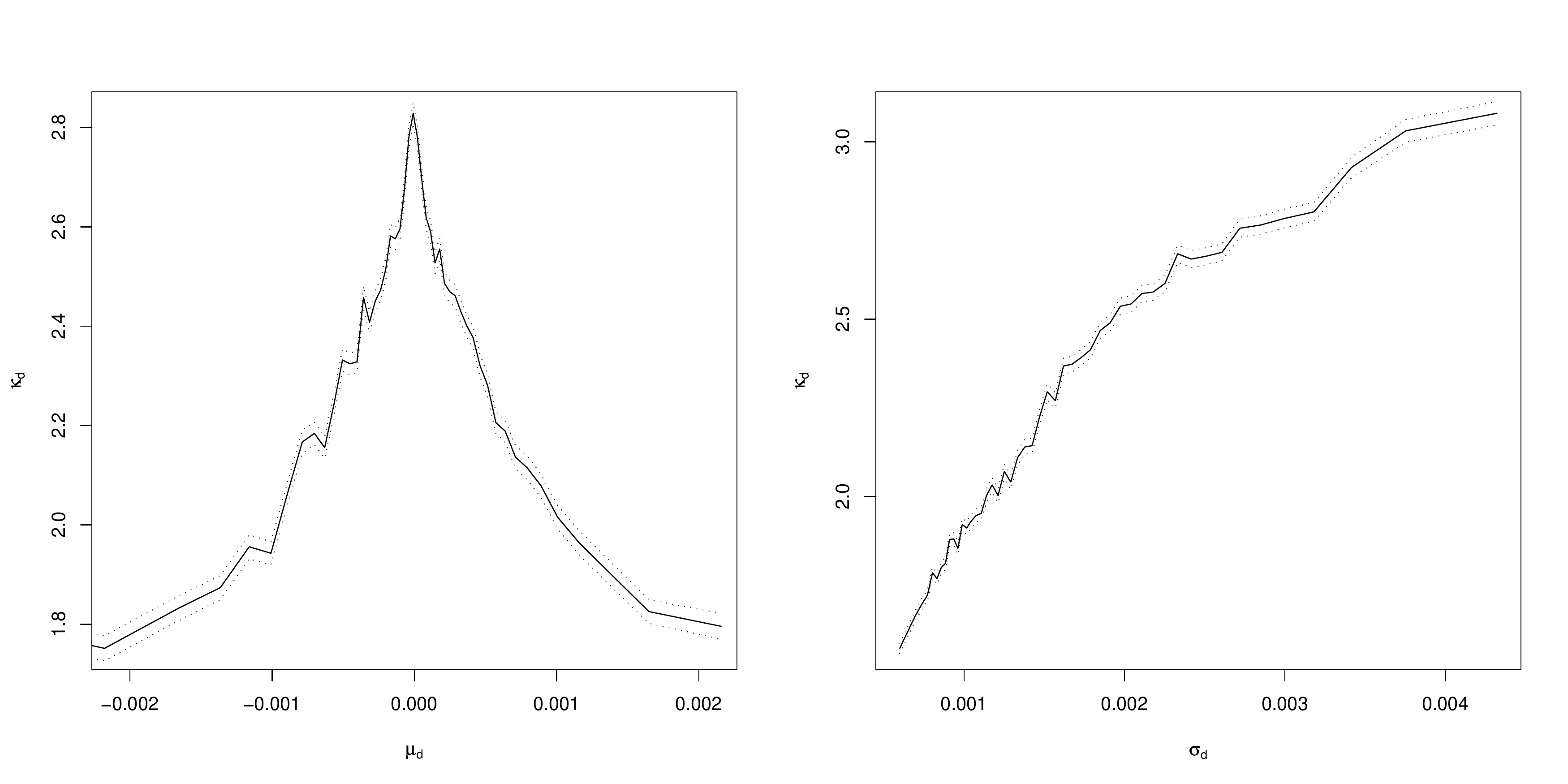}
		\caption{Left: Cross sectional kurtosis $\kappa_d$ as a function of the index return (equiweighted on all stocks) $\mu_d$. Note that the dependence is 
	        nearly the same for positive and negative market returns. Right: Cross sectional kurtosis $\kappa_d$ as a function of the cross sectional dispersion $\sigma_d$.
                We added error bars (dashed lines) on both figures.} 
        \label{figure5}
\end{figure}

Finally, the dispersion kurtosis $\kappa_d$ shows again a non-intuitive {\it decreasing} behaviour as a function of $|\mu_d|$, see Fig.~\ref{figure5}. The average
kurtosis conditioned to a value of $|\mu_d|$ decreases from $\approx 2.8$ for small index returns to $\approx 1.8$ for index returns larger than $2 \%$
in absolute value. This was first noticed {\it en passant} in \cite{Cizeau} on daily data and recently emphasised by Borland \cite{Lisa}. Here, we confirm on five minute returns
this strange stylised fact: the cross-sectional distribution of returns appears to be more Gaussian when its mean is off-centred. 

However, if we now condition $\kappa_d$ on the dispersion $\sigma_d$ (which, as we found above, is {\it positively} 
correlated with $|\mu_d|$), we
find (see Fig.~\ref{figure5}) the opposite behaviour, i.e. the larger the dispersion, the larger the kurtosis $\kappa_d$! 
We will offer a discussion of these confusing effects in the discussion section below.

\section{Intra-day seasonalities in the inter-stock correlations}

Let us now turn to the properties of the eigenvalues and eigenvectors of the $N \times N$ correlation matrix $C_{\alpha,\beta}(k)$ 
defined by Eq. (\ref{correlation}) above.

\subsection{The top eigenvalue}

The largest eigenvalue $\lambda_1$ of the correlation matrix of stock returns is well known to be associated with the ``market mode'', 
i.e. all stocks moving more or less in sync. As recalled above, the quantity $\lambda_1/N$ can be used to define the average correlation
between stocks. 

\begin{figure} 
		\center
		\includegraphics[scale=0.45]{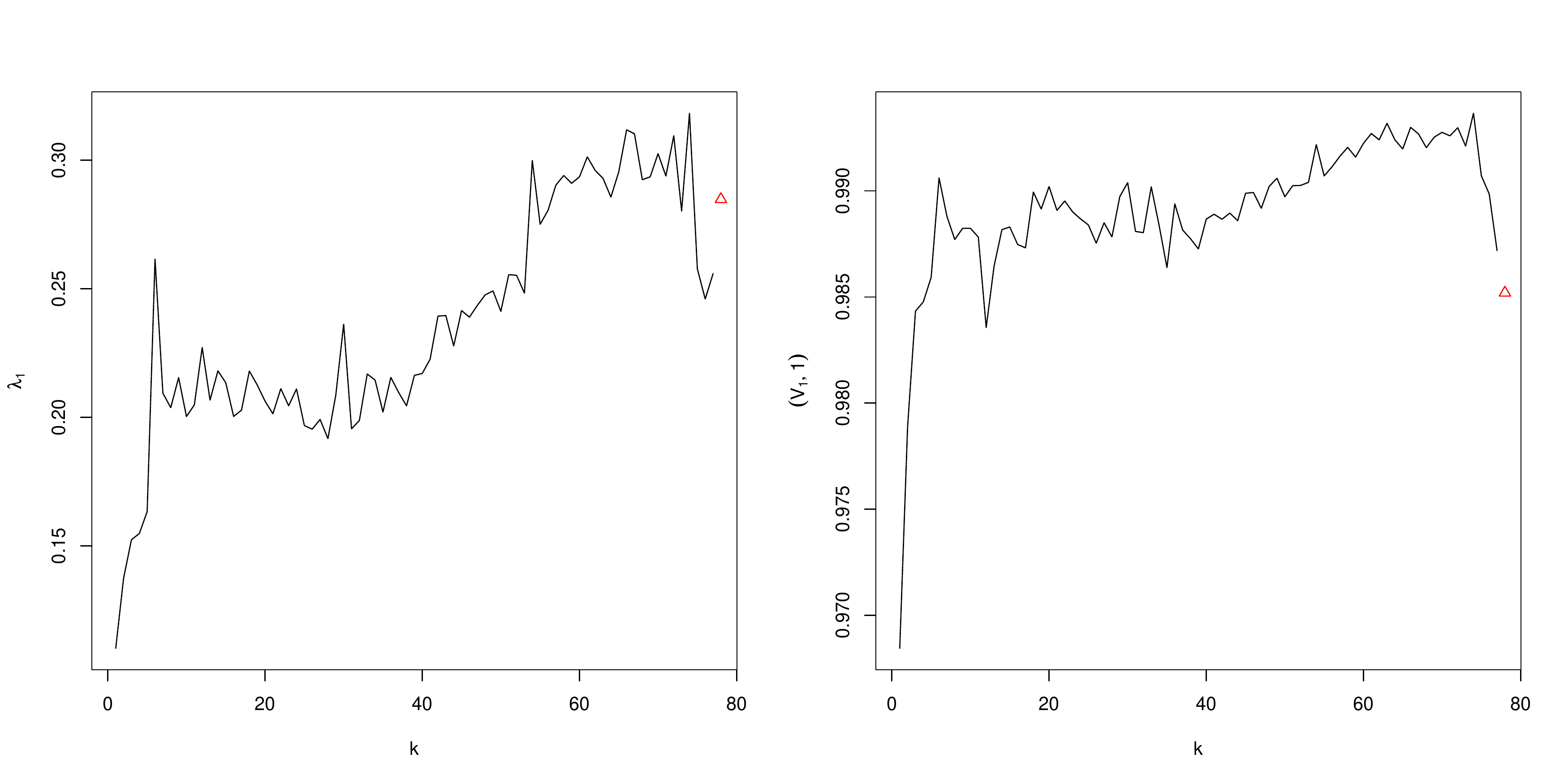}
		\caption{Left: Dependence of the top eigenvalue of the correlation matrix $C(k)$, $\lambda_1(k)/N$, as a function of time of day. Right: Evolution of the scalar product between 
		top eigenvector $\vec v_1(k)$ and the uniform vector $\vec e$. The red triangles correspond to the overnight values.} 
        \label{figure6}
\end{figure}

We show in Fig.~\ref{figure6} (left side) the magnitude of $\lambda_1/N$ as a function of $k$. Interestingly, the average correlation clearly increases as time elapses, 
from a rather small value $\approx 0.12$ when the market opens to $\approx 0.3$ near market close. This is in agreement with the fact that the
dispersion $\sigma_d(k)$ is, in relative terms, smaller at the end of the day (see Fig.~\ref{figure1}, inset). The value of $\lambda_1/N$ for the correlation
of overnight returns is also around $0.3$, in continuity with the value at the end of the trading day. 

In agreement with the idea that the stock
dynamics become more and more uniform as the day proceeds, we find a substantial increase of the scalar product of the largest eigenvector $\vec v_1(k)$ with the 
uniform normalised vector $\vec e=(1/\sqrt{N},1/\sqrt{N}, \dots, 1/\sqrt{N})$ --- see Fig.~\ref{figure6} right. This scalar product is always close to unity, 
confirming the market mode interpretation of the top eigenvalue, but starts the day around $0.97$ and ends the day at $0.995$, before dropping 
again in the last bins of the day and during the overnight, when it is equal to $0.985$ (i.e. larger than the open value).

\subsection{Smaller eigenvalues}

\begin{figure} 
		\center
		\includegraphics[scale=0.45]{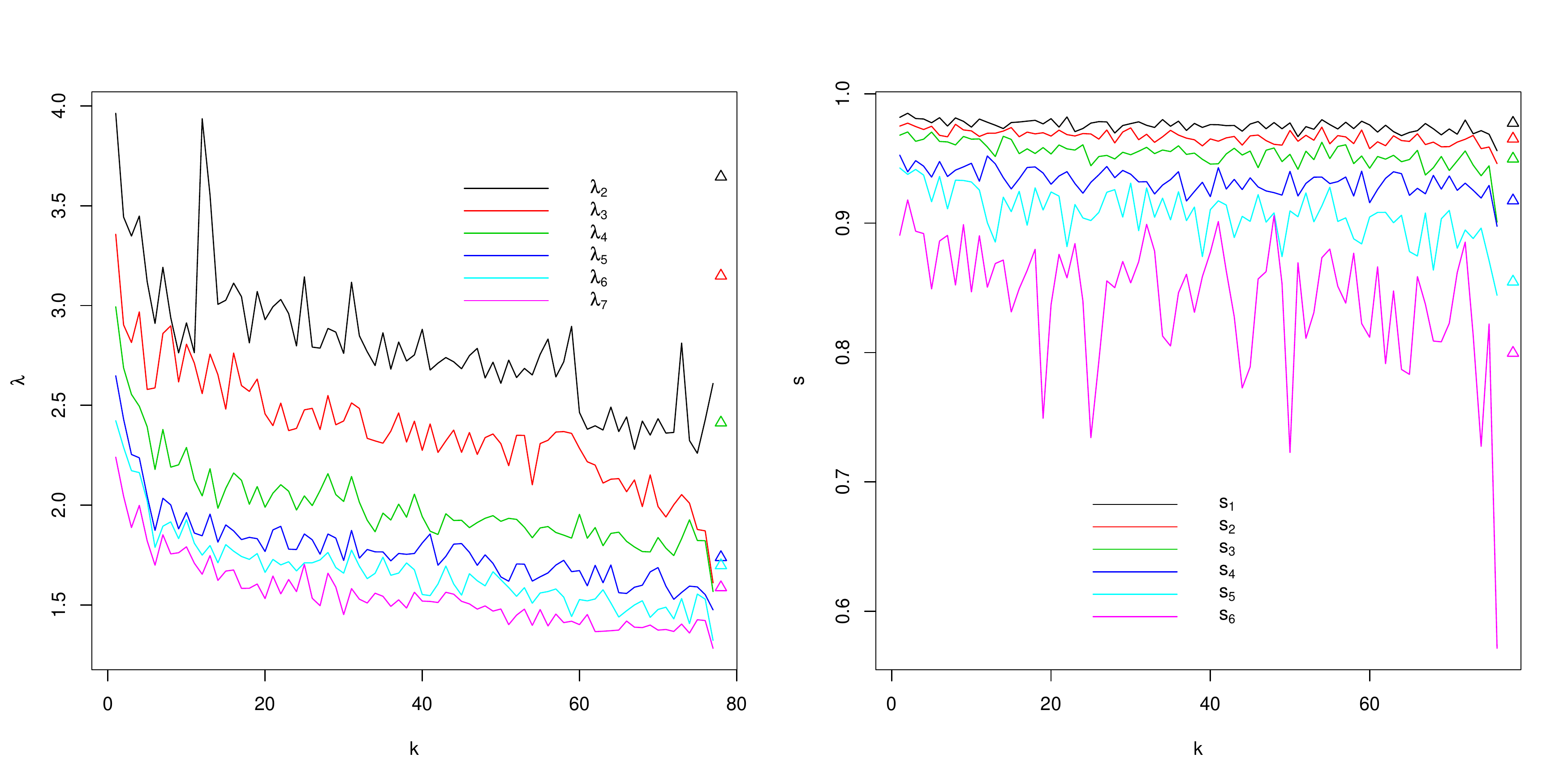}
		\caption{Left: Smaller eigenvalues $\lambda_i(k)$, $i=2,\dots,7$) as a function of $k$. Right: The 6 singular values $s_\ell(k)$ ($\ell=2,\dots,7$) 
		of the matrix $W_{ij}(k)$. In the absence of any true persistence, one would expect all singular value to lie in the interval $[0,0.12]$, much below the 
		smallest singular value $s_7 \sim 0.8$. The triangles correspond to the overnight values. } 
        \label{figure7}
\end{figure}

The evolution of the next six eigenvalues $\lambda_i(k)$, $i=2, \dots, 7$ is shown in Fig.~\ref{figure7}. We see that the amplitude of this risk factors now {\it decreases} 
with time, before shooting back up during the overnight (see the last point of the graphs). Although by construction the trace of the correlation matrix, and 
therefore the sum of all $N$ eigenvalues is constant (and equal to $N$), this decrease is not a trivial consequence of the increase of $\lambda_1$, since the 
sum of the first five eigenvalues is $\sim 50$, still small compared to $\operatorname{Tr}(C)=N=126$. What we see here is that as the day proceeds, more and more risk is carried 
by the market factor, while the amplitude of sectorial moves shrivels in relative terms (but remember that the correlation matrix is defined after normalising the 
returns by the local volatility, which increases in the last hours of the day).

It is more difficult to visualise the evolution of the corresponding eigenvectors, since there is no natural vectors to compare them with. Furthermore, eigenvalues can
``collide'' and cross, resulting in an interchange between two consecutive eigenvectors. We have therefore chosen to take as a reference the eigenvectors $\vec v_i(1)$ in the opening
bin $k=1$, corresponding to the largest values of $\lambda_i(k)$, $i=2, \dots, 7$. We then form the $6 \times 6$ matrix of scalar products $W_{ij}(k) = \vec v_i(1) \cdot \vec v_j(k)$. 
The singular values $s_\ell(k)$ of this matrix (equal to the square-root of the eigenvalues of $W^T \, W$) give a measure of the overlap between the eigenspace spanned by the $\vec v_i(1)$ 
and that spanned by the $\vec v_j(k)$. If the $\vec v_j(k)$ are a permutation of the $\vec v_i(1)$, all the $s_i$'s are equal to unity, indicating maximum overlap. 
In particular, $s_\ell(1) \equiv 1$ trivially. The evolution of the $s_\ell(k)$, $\ell=2, \dots, 7$ is shown in Fig.~\ref{figure7} right. Using the results of \cite{BLMC}, we conclude that 
all $s_\ell(k)$ are meaningful, since in the absence of any true correlations between the $\vec v_i(1)$ and the $\vec v_j(k)$, one would expect all singular values to lie in the interval $[0,0.12]$. 
Therefore, although the structure of correlations clearly evolves between the opening hours and the closing hours, there is as expected a strong overlap between the principal components 
throughout the day.

\section{Discussion \& Conclusion}

Let us present a synthetic account of the above empirical results, for which we only propose an interpretation stub. We have seen that during the opening hours of the market, the
volatility and the dispersion of returns are high, whereas kurtosis effects are relatively low. These two quantities are different measures of the {\it heterogeneity} of stock returns, and 
quite paradoxically they are found to behave in opposite ways. But while the volatility and dispersion are dimensional measures of heterogeneity (measuring the spread of returns in $\%$), the kurtosis 
is a relative, a-dimensional measure of {\it surprise}. What our results mean in intuitive terms is that although the typical amplitudes of stock returns are high in the morning, outliers are relatively rare, both over time 
and over stocks. In a sense, agitation is the norm during these early hours of trading, stocks move in different directions in such a way that the
average correlation is weaker than average, and the top eigenvector of the correlation matrix is farther away from the uniform mode  $\vec e=(1/\sqrt{N},1/\sqrt{N}, \dots, 1/\sqrt{N})$.
But anomalously large jumps rarely take place in the morning --- as expected, these jumps are more likely overnight (and are to be related to arrival of corporate specific or market-wide information), where kurtosis effects 
are strongest, both for single stock and cross-sectional returns. As the day proceeds, 
correlations increase and dispersion decreases, but {\it unexpectedly large jumps become more probable}, thereby increasing the kurtosis.

The second somewhat paradoxical effect is the dependence of the kurtosis on the index return, which was recently interpreted by L. Borland as a signature of collective behaviour during crises \cite{Lisa}.
Again, days when the market as a whole moves a lot are also large dispersion days where all stocks move a lot in different directions, but with little outliers, i.e. one or a handful of stocks that would 
jump up or down. In this sense, these days are more homogeneous. Should one deduce from this that there is a stronger ``synchronisation'', or collective dynamics, during these periods, as suggested by Borland? While it is 
true that the average correlation between stocks depends on the index return, this dependence is in fact signed: correlations are stronger for negative index returns and weaker for positive returns, see \cite{Ingve,PAskew}. 
This is in contrast with the kurtosis effect discussed here, which is surprisingly symmetrical 
(see Fig.~\ref{figure5}). A quantitative model for this behaviour is missing at this stage. 
Qualitatively, however, we believe that the mechanism is the following \cite{Remi}: when the index return is large, the dominant source of dispersion becomes the market exposure (the `$\beta$'s') of the different stocks, 
rather than the idiosyncratic residuals. Since the distribution of the $\beta$'s is roughly Gaussian, kurtosis effects do indeed decrease for large index returns. This interpretation however requires that the volatility of the residuals 
increases sub-linearly with the index volatility, as indeed suggested by the data shown in Fig.~\ref{figure3}. The fact that during large swings of the
index, the market exposure of stocks becomes the dominant factor is probably a result of index/futures arbitrage. 

Finally, although large index return days are large dispersion days, the converse is not true. 
A typical large dispersion day is in fact a day when one or a handful of stocks gyrate wildly,
contributing both to the dispersion and to the kurtosis, and explaining the {\it positive} correlation between $\sigma_d$ and $\kappa_d$. If this interpretation is correct, this positive correlation
should diminish when one uses the mean-absolute deviation and not the variance to compute the dispersion, since the former is less sensitive to outliers. We have checked that this is indeed the case.

To summarise, we have established several new stylised facts concerning the intra-day seasonalities of stock dynamics. Beyond the well known U-shaped
pattern of the volatility, we have found that the average correlation between stocks increases throughout the day, leading to a smaller dispersion
between stocks (in relative terms). However, the kurtosis, which is a measure of volatility surprises, is in fact minimum at the open of the market, when the volatility is at its peak. We have also confirmed that the dispersion kurtosis 
is a symmetric, markedly decreasing function of the index return. This
means that during large market swings, the idiosyncratic component of the stock dynamics becomes sub-dominant, an effect that we have confirmed directly. 
Finally, while the market mode component of the dynamics becomes stronger as the day proceeds, the sectorial components recede. In a nutshell, early 
hours of trading are dominated by idiosyncratic or sector specific effects with little surprises, whereas the influence of macro, market factor increases throughout the day, and surprises become more frequent. 
A detailed quantitative interpretation of our results, for example of the power-law decay of the volatility in the morning, is at this stage lacking. We believe that, when available, such an interpretation
will shed light on the relative importance of behavioural and informational effects on price formation and volatility.

\vskip 1cm
{\it Acknowledgements} We have benefitted from insightful comments and suggestions by R\'emy Chicheportiche, Stefano Ciliberti, Benoit Crespin, Laurent Laloux, Marc Potters, Pierre-Alain Reigneron 
and Vincent Vargas. Some of our results have been independently obtained by L. Borland (private communication and in preparation).

\bibliography{biblio_paper}
\bibliographystyle{plain}

\end{document}